\documentclass[aps,prd,preprint]{revtex4-1}
\usepackage[utf8]{inputenc}
\usepackage[pdftex]{graphicx}
\usepackage{amsmath,amssymb}
\usepackage{mathtools}
\usepackage{hyperref}
\hypersetup{
    colorlinks,
    citecolor=blue,
    filecolor=blue,
    linkcolor=blue,
    urlcolor=blue
}
\begin{document}

\title{The CSS Hamiltonian: high energy evolution of rapidity dependent observables}
\date{\today}
\author{Haowu Duan$^a$, Alex Kovner$^a$ and  Michael Lublinsky$^b$}

\affiliation{$^a$ Physics Department, University of Connecticut, Storrs, CT 06269, USA\\
$^b$ Physics Department, Ben-Gurion University of the Negev, Beer Sheva 84105, Israel}
\begin{abstract}
{We consider  evolution of observables which depend on a small but fixed value of longitudinal momentum fraction $x$, to high rapidity, such that $\eta>\ln 1/x$. We show that this evolution is not given by the JIMWLK (or BK) equation. We derive the evolution Hamiltonian - $H_{CSS-JIMWLK}$ which generates this evolution in the cases of dilute and dense projectile wave function. The two limits yield identical results for $H_{CSS-JIMWLK}$. We show that the resulting evolution for the gluon TMD is identical to the (double logarithmic) perturbative Collins-Soper-Sterman evolution equation in the longitudinal resolution parameter at a fixed and very large transverse resolution.}
\end{abstract}

\maketitle
\section{Introduction}
The study of high energy evolution of hadronic observables has been a very lively and active research area in the QCD community. The evolution of hadronic cross sections with energy has been  formulated in several different ways, which are fundamentally equivalent to each other: the Balitsky hierarchy (or BK equation in its mean field incarnation) for correlators of Wilson lines, or the JIMWLK functional evolution  equation for the color field quasi probability density functional\cite{Balitsky:1995ub,JIMWLK}.
At leading order in $\alpha_s$ these have been known for over 25 years, and their NLO form has also been derived and studied more recently \cite{NLOJIMWLK}.

In all of these approaches the object of interest is a set of hadronic observables dominated by gluons (or quarks) with lowest allowed longitudinal momenta present in the hadronic state boosted to high rapidity. However these do not exhaust  all interesting observables. 

In particular lately there has been a lot of interest of understanding the interplay of the Sudakov suppression and high energy evolution in processes with two very different transverse scales at high energy. One example of such process considered recently by several groups is production of high $p_\perp$ dijets with small momentum imbalance in photon (real or virtual) hadron scattering at high energy \cite{dijet}.
The Sudakov suppression in such a process is expected. Usually such a suppression can be understood as either gluon radiation attributable to the "hard part" of the process, in this case emi=ssion from the dijet, or alternatively as a result of the Collins, Soper, Sterman (CSS) evolution of the target transverse momentum distribution (TMD) \cite{CSS}. In order to understand it as latter, one needs to be able to calculate gluon TMD at fixed longitudinal momentum fraction $x$ in the target wave function which is evolved well past this value of $x$. This is not an observable of the type described by JIMWLK evolution, since the gluons one is tagging are not the softest gluons in the evolved wave function. We will call such observables, observables of the CSS type.

Another example of an observable where not only the softest gluons are important is multi gluon production cross section where the rapidity differences between the tagged gluons are large \cite{Jalilian-Marian:2004vhw,Kovner:2006wr}. In this case one has to evolve the hadronic wave function in rapidity well past the rapidities of some of the gluons, and this part of the evolution is not going to be JIWMLK like.

In the present paper we study the evolution of observables of this type. We derive the analog of the JIMWLK Hamiltonian, which has to be used for evolving observables of non-JIMWLK type, in the sense that they measure properties of gluons at fixed, small  value of $x$ which is nevertheless 
greater than  $x$ of the softest gluons in the evolved wave function. We call this Hamiltonian - the CSS-JIMWLK Hamiltonian, $H_{CSS-JIMWLK}$. 

We start in Sec. 2 by reviewing the derivation of the standard JIMWLK Hamiltonian. Several such derivations exist in the literature using different perspectives \cite{Balitsky:1995ub,JIMWLK,Mueller:2001uk,ming}. Here we work with the derivation presented in a way convenient for the generalization/modification we are seeking \cite{ming}.  
In Sec. 3 we explain how this derivation has to be modified if observables of the CSS-type are to be considered and derive $H_{CSS-JIMWLK}$ both in the dense and dilute projectile limits. Perhaps surprisingly, in these two limits the Hamiltonian turns out to be the same when written in terms of the color charge currents of the projectile. In Sec. 4 we  show that when applied to the gluon (and quark) TMD the evolution generated by $H_{CSS-JIMWLK}$ yields the CSS equation with respect to longitudinal resolution with transverse resolution equal to UV scale. Finally Sec.5 contains a short discussion of our results.

\section{The JIMWLK evolution: a recap}

We start with a recap of the derivation of the JIMWLK evolution equation. In this we mostly follow the exposition in \cite{ming}. Our aim here is not to repeat the derivation of JIMWLK for the n-th time, but rather to set the stage for the subsequent discussion of the CSS-JIMWLK Hamiltonian.

\begin{figure}[t]                                       
                                  \includegraphics[width=7cm] { 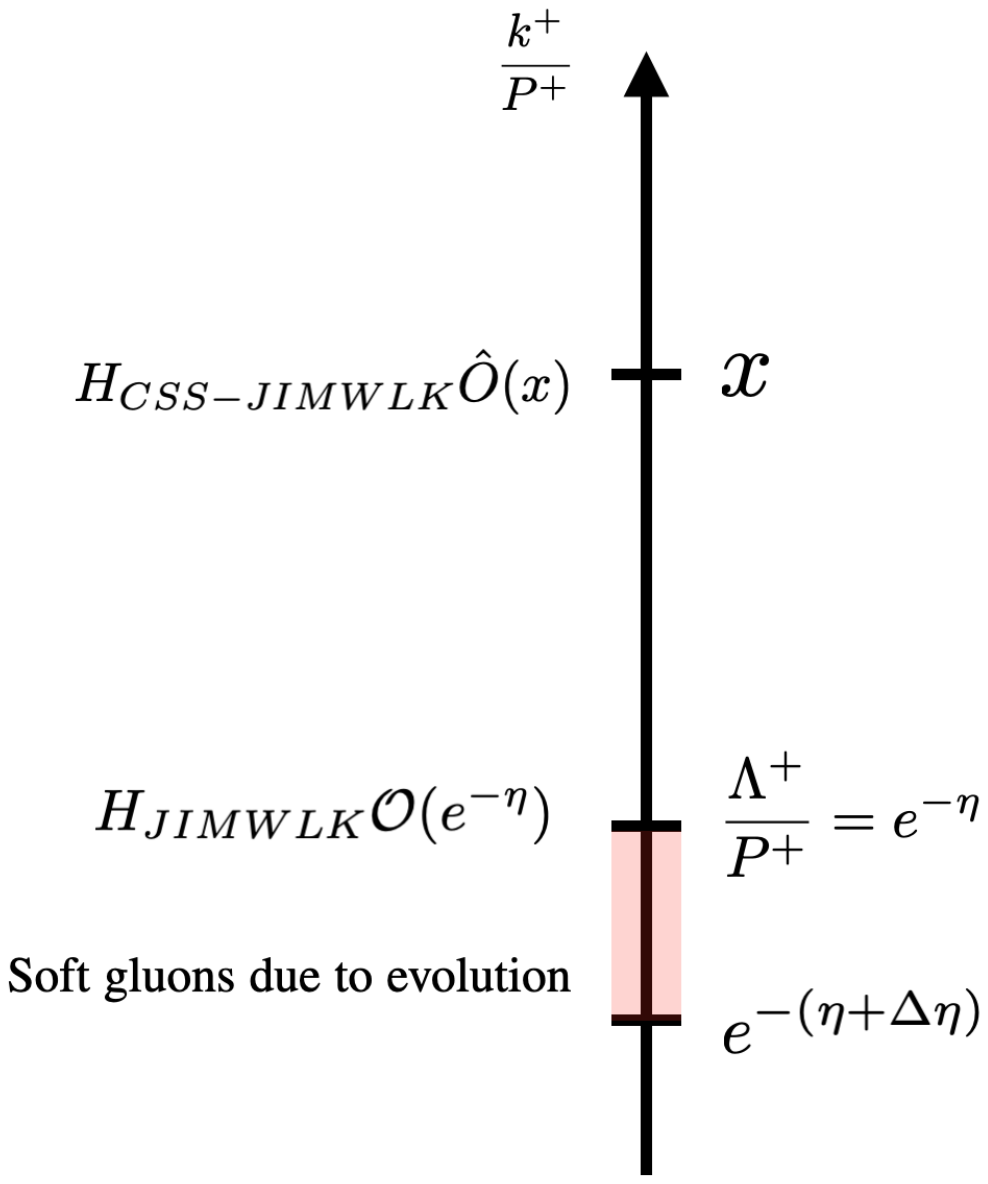}             
\caption{The mode structure of the wave function and the two types of operators/evolutions.}
\label{f0}
\end{figure}

Consider a wave function of a hadronic system, which we will refer to as "the projectile". We separate the degrees of freedom of this system into "valence" and "soft", see Fig. \ref{f0}. The valence degrees of freedom are the gluon fields with longitudinal momentum greater than some cutoff value $\Lambda^+$, or equivalently longitudinal momentum fraction greater than  $ \Lambda^+/P^+\equiv e^{-\eta}$, 
where $P^+$ is the total longitudinal momentum of the hadron and $\eta$ is the maximal rapidity of the gluons in the hadron.   
The soft degrees of freedom are gluons with smaller longitudinal momentum fraction. They occupy a window of rapidity $\Delta\eta$ below 
the cutoff $\Lambda^+$.

At initial energy the valence gluons are assumed to scatter eikonally on the target, while the soft gluons do not contribute to scattering. If the target is dense, one takes into account multiple "soft" scatterings, if it is dilute the scattering is considered to leading order only. To increase the energy of the collision we boost the hadronic projectile. As a result, part of the soft gluons become energetic enough to scatter, and this leads to the evolution of the scattering amplitude.

The valence system evolved to rapidity $\eta$ is described by the valence state denoted as $|\eta\rangle$. The wave function of the soft gluons in the presence of the valence charges has the form of a unitary operator $\Omega$  acting on the Fock space vacuum $|0\rangle_s$
of the soft modes, so that the complete wave function is
\begin{equation}\label{wf}
\Psi=\Omega[\hat j, a^\dagger,a]|0\rangle_s\otimes|\eta\rangle\,;~~~~~~~~~ 
\hat j^a(\mathbf{x}_{\perp})\,=\,\int_{\Lambda^+}^\infty {dk^+\over 2\pi}\,
a_v^\dagger(k^+,\mathbf{x}_{\perp})\,T^a\,a_v(k^+,\mathbf{x}_{\perp})
\end{equation}
where $a^\dagger$ and $a$ are creation and annihilation of the soft gluon modes, while $a_v^\dagger$ and $a_v$ refer to the valence gluon creation and annihilation operators.
The unitary operator $\Omega$ has been calculated for dilute projectile, where the color current parametrically is considered to be $j\sim 1$, and for dense case where $j\sim 1/\alpha_s$. For dilute projectile, $\Omega$ is a simple coherent operator 
 \begin{equation}\label{eq:coherent_operator}
 \Omega = \mathrm{Exp}\left\{i\int d^2\mathbf{x}_{\perp}\,\hat b_i^a(\mathbf{x}_{\perp}) 
 \int_{\Lambda^+e^{-\Delta \eta}}^{\Lambda^+} \frac{dk^+}{\pi |k^+|^{1/2}}\left({a}_i^{\dagger a}(k^+, \mathbf{x}_{\perp}) + {a}_i^{a}(k^+, \mathbf{x}_{\perp})\right)\right\}
 \end{equation}
where the longitudinal integral is over the 
%newly opened 
"window" of rapidity width $\Delta \eta$, and $\hat b^a_i(x_\perp)$ is the "classical" field created by the color charge $\hat j$. 

 \begin{equation}\label{dilute}
 \hat b_i^a(\mathbf{x}_{\perp})=g\int d^2\mathbf{y_\perp} \frac{\partial_i}{\partial^2}(\mathbf{x_\perp,y_\perp})\,\hat j^a(\mathbf{y}_\perp)
 \end{equation}
For dense projectile the field is given by the solution of classical Yang-Mills equations
 \begin{equation}\label{bj}
 \begin{split}
& \partial_i \hat b_i^a(\mathbf{x}_{\perp}) =g \hat j^a(\mathbf{x}_{\perp})\, ,\\
&\partial_i \hat b_j^a(\mathbf{x}_{\perp}) - \partial_j \hat b_i^a(\mathbf{x}_{\perp}) -gf^{abc} \hat b_i^b(\mathbf{x}_{\perp}) \hat b_j^c(\mathbf{x}_{\perp}) =0 \, .\\
\end{split}
 \end{equation}
In the dense case, $\Omega$ is not a simple coherent operator, but in addition contains a factor  Gaussian in the creation and annihilation operators \cite{Kovner:2007zu,tolga}. The explicit form of $\Omega$ is not important here, but an interested reader can find the derivation in \cite{tolga}.

%If 
Prior to  the evolution by a rapidity $\Delta \eta$ the "valence" system is described by a density matrix 
$\hat\rho_v=|\eta\rangle\langle\eta|$ (the derivation is equally valid for a valence state which is a mixed state).
For  observables $\hat O$ that depend on the total valence charge density $\hat j$, 
\begin{equation}
 {\rm Tr}[\hat O(\hat j)\hat\rho_v]\,=\,\int dj\,\tilde O(j)\,W_\eta[j]
\end{equation}
Here the expectation value of $\hat O$ has been represented using the Wigner-Weyl formulation of quantum mechanics, introducing
the phase space (quasi) probability distribution $W_\eta[j]$ of the classical phase space variable $j$.  The function $\tilde O(j)$ is the 
Wigner-Weyl  representation of the the operator $\hat O$ (see 
\cite{ming},\cite{Kovner:2005uw}).

After one step of the evolution the density matrix is
\begin{equation}\label{deltarho}
 \hat{\rho}(\Delta \eta) \,=   \,{\Omega} |0\rangle \hat{\rho}_v \langle 0|\,{\Omega}^{\dagger}\, .
 \end{equation}

The JIMWLK equation is derived for a set of observables that depend on the total valence charge density, i.e. charge density integrated over rapidity (see eq.(\ref{wf})). As such at one step of the evolution not only the wave function of the system evolves, but also the operator of the observable gets an additional contribution due to the color charge density in the new rapidity interval that opens up due to the evolution. The evolution of such an observable follows from the calculation of the expectation value 
 \begin{equation}\label{ohat}
 {\rm Tr}[\hat O(\hat j+\hat j_{\mathrm{soft}})\,\hat\rho(\Delta \eta)]={\rm Tr}[\hat{R}^{\dagger}\,\hat O( \hat{j})\,\hat{R} \,\hat\rho(\Delta \eta)]={\rm Tr}[\hat O( \hat{j})\,\hat{R} \,\hat\rho(\Delta \eta)\,\hat{R}^{\dagger}]
 \end{equation}
where $\hat\rho(\Delta \eta)$ is given in \eqref{deltarho}, and the operator $\hat R$ shifts the valence color charge density according to
 \begin{equation}\label{rj}
 \hat{R}^{\dagger} \hat{j}^a(\mathbf{x}_{\perp})\hat{R} \,=\, \hat{j}^a(\mathbf{x}_{\perp}) + \hat{j}^a_{\mathrm{soft}}(\mathbf{x}_{\perp})\, ;
 ~~~~~~~
 \hat j^a_{soft}(\mathbf{x}_{\perp})\,=\,\int^{\Lambda^+}_{\Lambda^+ e^{-\Delta\eta}} {dk^+\over 2\pi}\,
 a^\dagger(k^+,\mathbf{x}_{\perp})\,T^a\,a(k^+,\mathbf{x}_{\perp}).
 \end{equation}
 That such an operator can be formally mathematically defined was shown in \cite{ming}.
The evolution can be conveniently represented 
%in the Wigner-Weyl formulation of quantum mechanics, 
as the evolution of the 
%phase space (quasi) 
probability distribution $W_\eta[j]$. 
% of the classical phase space variable $j$ (see 
%\cite{Kovner:2005uw,Altinoluk:2013rua}). 

The evolution equation follows from performing the average in \eqref{ohat} over the soft gluon Hilbert space:
\begin{equation}
\frac{d W_\eta[j^a]}{d\eta}\,  =\, -\,H_{RFT}\,W_\eta[j^a]
\end{equation}
with the "master Hamiltonian"
\begin{equation}\label{wj}
H_{RFT}=
\frac{1}{4\pi} \Bigg[(\tilde{b}_L^{a} - b_R^{a})^{\dagger}(N^{\dagger}_{\perp,R}N_{\perp,R} )^{ab}(\tilde{b}_L^{b} -b_R^{b}) +h.c.\Bigg]
%H_{RFT}=
%\frac{1}{4\pi} \Bigg[(\tilde{b}_L^{\alpha} - b_R^{\alpha})^{\dagger}(N^{\dagger}_{\perp,R}N_{\perp,R} )_{\alpha\beta}(\tilde{b}_L^{\beta} -b_R^{\beta}) +h.c.\Bigg]
\end{equation}
where the summation over color indices and integration over transverse coordinates is implicit.

The Hamiltonian in \eqref{wj} is written in terms of classical fields $b$ defined as solutions of classical Yang-Mills equations with the source $j$
which formally look identical to (\ref{bj}), except that the fields and charges are not operators in the Hilbert space, but functions on the classical phase space.
The factor $N_\perp$ is given by
\begin{equation}\label{eq:Nperp_exp1}
N_{\perp} [b]= 
%[1-l-L] = 
\left[\delta_{ij} - \partial_i \frac{1}{\partial^2}\partial_j -D_i\frac{1}{D^2} D_j\right]^{ab} (\mathbf{x}_{\perp}, \mathbf{y}_{\perp})\, ;
~~~~~~~~~D_i^{ab} = \delta^{ab}\partial_i -igT^{e}_{ba}b_i^e 
\end{equation}
In \eqref{wj},
 \begin{equation} \label{definitions}
 b_L^{b}\equiv b^{b}[j_L]; ~~~~~ b_R^{b}\equiv b^{b}[j_R]; ~~~~~~N_{\perp,R}=N_\perp[b_R].\,;~~~~~~
 \tilde{b}_L^{a} = b_L^{b}\,{\cal R}_{p}^{ab}\,;~~~~~~~~~{\cal R}_{p}\,=\,e^{\tau}\,;   ~~~~~
 \tau = gT^e\frac{\delta}{\delta j^e}
 \end{equation}
The left and right currents $j_{L(R)}$ are expressed in terms of the classical phase space variable $j$ as 
\cite{Kovner:2005uw,Altinoluk:2013rua} 
\begin{equation}
\begin{split}
j_L^a =&j^b\left[\frac{\tau}{2}\coth{\frac{\tau}{2}} + \frac{\tau}{2}\right]^{ba}\\
= &j^a + \frac{1}{2} j^b \left(gT^e\frac{\delta}{\delta j^e}\right)_{ba}+\frac{1}{12} j^b \left(gT^e\frac{\delta}{\delta j^e}\right)^2_{ba} -\frac{1}{720}j^b\left(gT^e\frac{\delta}{\delta j^e}\right)^4_{ba} +\ldots\\
j_R^a  =&j^b\left[\frac{\tau}{2}\coth{\frac{\tau}{2}} - \frac{\tau}{2}\right]^{ba}\\
 =& j^a - \frac{1}{2} j^b \left(gT^e\frac{\delta}{\delta j^e} \right)_{ba}+\frac{1}{12} j^b \left(gT^e\frac{\delta}{\delta j^e}\right)^2_{ba} -\frac{1}{720}j^b\left(gT^e\frac{\delta}{\delta j^e}\right)^4_{ba}+ \ldots\\
\end{split}
\end{equation}
%with 
%\begin{equation}
%\tau = gT^e\frac{\delta}{\delta j^e}
%\end{equation}
The introduction of distinct left and right operators is an integral part of representing quantum mechanics as a theory in the classical phase space \cite{Agarwal:1971wc}.
Although $\tau$ is a differential operator, in the above expressions it is treated as a number. This is because at the end of the day it acts only on $W_\eta[j]$ and does not act on any of the factors $j$ in the evolution Hamiltonian (\ref{wj}) itself.
%Also in \eqref{wj},   
%$\tilde{b}_L^{a} = b_L^{b}\,{\cal R}_{p}^{ab}$ with
%\begin{equation}
%\tilde{b}_L^{a} = b_L^{b}\,{\cal R}_{p}^{ab}\,;~~~~~~~~~ {\cal R}_{p}\,=\,e^{gT^a\frac{\delta}{\delta j^a}}\,=\,e^{\tau}
%\end{equation}
Also  ${\cal R}_p$ is treated as a matrix in color indices and not a differential operator.

A physically intuitive way of thinking about the classical derivative operator $\tau$ is that it represents the color fields of the target on which the projectile is scattering. Formally the derivation does not assume scattering at all, but deals only with the projectile wave function. However if we consider the  scattering matrix of the projectile on a target described by the color fields $\alpha_T$, in the eikonal approximation, every factor of $\tau$ turns into a factor of $\alpha_T$, since the calculation of the S-matrix  is equivalent to functional Fourier transform \cite{Kovner:2005en}.

 Although 
Eq.\eqref{wj} is written in the form which seemingly can be applied to the situation where the target and projectile can be both dense,  this is not quite the case. 
The derivation of Eq.\eqref{wj} is fully under control only in the situation where at least one of the scattering objects (target and projectile) is dilute. The general form given in Eq.\eqref{wj} should be understood as an interpolating function between these limiting situations. The two physical limits of Eq.\eqref{wj} are the so called JIMWLK and KLWMIJ Hamiltonians obtained in the limit of dilute target and dilute projectile respectively.

The JIMWLK Hamiltonian applies to a dense projectile scattering on a dilute target. To obtain the JIMWLK Hamiltonian one expands Eq.\eqref{wj} to leading (second) order in $\tau$.  At this order we have $N_{\perp,R(L)}=N_\perp$
and
\begin{equation}
\begin{split}
&b_L{\cal R}_p - b_R  = b_i^a[j_L] {\cal R}_p^{ab} - b_i^b[j_R] \\
\simeq & \left(b_i^a[j] + \frac{\delta b_i^a}{\delta j^e}(j_L^e -j^e)\right)\left(\delta_{ab}+ gT^d_{ab}\frac{\delta}{\delta j^d}\right) - \left(b_i^b[j] +\frac{\delta b_i^b}{\delta j^e} (j_R^e -j^e)\right)\\
\simeq&gb_i^aT^d_{ab}\frac{\delta}{\delta j^d} + \frac{\delta b_i^b}{\delta j^e}g j^cT^d_{ce}\frac{\delta}{\delta j^d}\\
=&gb_i^a(\mathbf{x}_{\perp})T^d_{ab}\frac{\delta}{\delta j^d(\mathbf{x}_{\perp})} + \int d^2\mathbf{z}_{\perp}\frac{\delta b_i^b(\mathbf{x}_{\perp})}{\delta j^e(\mathbf{z}_{\perp})} gj^c(\mathbf{z}_{\perp})T^d_{ce}\frac{\delta}{\delta j^d(\mathbf{x}_{\perp})}\\
=&i\left[\partial_i - D_i\frac{1}{\partial D} D\partial \right]^{bd}(\mathbf{x},\mathbf{z})\, \frac{\delta}{\delta j^d(\mathbf{z}_{\perp})}
\end{split}
\end{equation}
where the covariant derivative is defined  in (\ref{definitions}).
%as $D_i^{ab} = \delta^{ab}\partial_i -igT^{e}_{ba}b_i^e$.
 In the last line we have used $igT^{e}_{ba}b_i^e = \delta^{ab}\partial_i - D_i^{ab} $ 
and $
igT^e_{ba}j^e = -(\partial D - D\partial)^{ab} = igj^eT^a_{eb} $ as well as 
\begin{equation}
\frac{\delta b_i^b(\mathbf{x}_{\perp})}{\delta j^e(\mathbf{z}_{\perp})} = \left[D_i\frac{1}{\partial D}\right]^{be}(\mathbf{x}_{\perp} ,\mathbf{z}_{\perp})\, 
\end{equation}
which follows by differentiating the equations \eqref{bj}.

Additionally using eq. \eqref{eq:Nperp_exp1} we calculate
\begin{equation}\label{eq:BL_BR1}
Q_i^a({\mathbf x}_\perp;U)\equiv [UN_{\perp} (b_L\mathcal{R}_p -b_R)]^a_i({\mathbf x}_\perp)=-i\left[U({\mathbf x})\left(D_i\frac{1}{D^2} - \partial_i\frac{1}{\partial^2}\right)D\partial\right]^{ab}(\mathbf{x}_\perp,\mathbf{z}_\perp) \frac{\delta }{\delta j^b(\mathbf{z}_\perp)}\, .
\end{equation}
where the standard eikonal scattering matrix $U$ is related to the classical field $b$ via
\begin{equation}
U(\mathbf{x}_{\perp}) = \mathcal{P} \mathrm{exp}\left[ ig\int_{\mathcal{C}} d\mathbf{y}_{\perp} \cdot b^c(\mathbf{y}_{\perp})T^c \right]
\end{equation}
with the path $\mathcal{C}$ starting at a fixed reference point (which can be taken at infinity) on the transverse plane and ending at the point $\mathbf{x}_{\perp}$. 

It is customary to represent the JIMWLK evolution in terms of the eikonal matrix $U$ as well as the left and right rotation operators that act on it. To do that one has to change variables from $j$ to $U$ \cite{ming},
\begin{equation}
\frac{\delta}{\delta j^a(\mathbf{x}_{\perp})} = \int d^2\mathbf{z}_{\perp} \frac{\delta U^{cd}(\mathbf{z}_{\perp})}{\delta j^a(\mathbf{x}_{\perp})} \frac{\delta}{\delta U^{cd}(\mathbf{z}_{\perp})}
\end{equation}
with
\begin{equation}
\begin{split}
 \frac{\delta U^{cd}(\mathbf{z}_{\perp})}{\delta j^a(\mathbf{x}_{\perp})} &=\int d^2\mathbf{y}_{\perp}  \frac{\delta U^{cd}(\mathbf{z}_{\perp})}{\delta b_l^b(\mathbf{y}_{\perp})}\frac{\delta b_l^b(\mathbf{y}_{\perp})}{\delta j^a(\mathbf{x}_{\perp})} 
% & =ig \int_{\mathcal{C}} d^2 y_l \left(U^{\dagger bm}(\mathbf{y}_{\perp})\, T^{m}_{cn}\right) U^{nd}(\mathbf{z}_{\perp}) \left[D_l\frac{1}{\partial D}\right]^{ba}(\mathbf{y}_{\perp} , \mathbf{x}_{\perp})\\
 %& = ig \int_{\mathcal{C}} d^2 y_l \left(U^{\dagger bm}(\mathbf{y}_{\perp})\, T^{m}_{cn}\right) U^{nd}(\mathbf{z}_{\perp}) \left[U^{\dagger} \partial_l U\frac{1}{\partial D}\right]^{ba}(\mathbf{y}_{\perp}  , \mathbf{x}_{\perp})\\
 %&= igT^{e}_{cn}U^{nd}(\mathbf{z}_{\perp})  \int_{\mathcal{C}} d^2 y_l  \left[ \partial_l U\frac{1}{\partial D}\right]^{ea}(\mathbf{y}_{\perp}  , \mathbf{x}_{\perp})\\
 %&= igT^{e}_{cn}U^{nd}(\mathbf{z}_{\perp})  U^{eb}(\mathbf{z}_{\perp})\left[ \frac{1}{\partial D}\right]^{ba}(\mathbf{z}_{\perp} ,\mathbf{x}_{\perp}) \\
 %& = ig U^{cm}(\mathbf{z}_{\perp}) T^{m}_{db} \left[ \frac{1}{\partial D}\right]^{ba}(\mathbf{z}_{\perp}, \mathbf{x}_{\perp}) \\
 \, =\,-\, ig\left[ U(\mathbf{z}_{\perp}) T^d \frac{1}{\partial D} \right]^{ca}(\mathbf{z}_{\perp} , \mathbf{x}_{\perp})
 \end{split}
\end{equation}
%In the above we have used  $U^{\dagger} T^a U = U^{ab}T^b$ and $D_l = U^{\dagger}\partial_l U$. %and also perform the contour integration by replacing the integral with the boundary values at $\mathbf{z}_{\perp}$. 
%We have also used 
%\begin{equation}
%\begin{split}
%\frac{\delta U^{cd}(\mathbf{z}_{\perp})}{\delta b_l^b(\mathbf{y}_{\perp})} =& U^{ce}(\mathbf{y}_{\perp})( ig T^b_{ef} )\int_{\mathcal{C}} d\mathbf{w}_l \delta(\mathbf{y}_{\perp} - \mathbf{w}_{\perp})[U^{\dagger}(\mathbf{y}_{\perp}) U(\mathbf{z}_{\perp})]^{fd}\\
%=&ig \left(U^{\dagger bm}(\mathbf{y}_{\perp})\, T^{m}_{cn}\right) U^{nd}(\mathbf{z}_{\perp})\int_{\mathcal{C}} d\mathbf{w}_l \delta(\mathbf{y}_{\perp} - \mathbf{w}_{\perp})
%\end{split}
%\end{equation}
This yields
\begin{equation}\label{dj}
%\begin{split}
\frac{\delta}{\delta j^a(\mathbf{x}_{\perp})} = \int d^2\mathbf{z}_{\perp} \frac{\delta U^{cd}(\mathbf{z}_{\perp})}{\delta j^a(\mathbf{x}_{\perp})} \frac{\delta}{\delta U^{cd}(\mathbf{z}_{\perp})}\
%&=ig\int d\mathbf{z}_{\perp}U^{cm}(\mathbf{z}_{\perp}) T^{m}_{db} \left[ \frac{1}{\partial D}\right]^{ba}(\mathbf{z}_{\perp}, \mathbf{x}_{\perp})\frac{\delta}{\delta U^{cd}(\mathbf{z}_{\perp})}\\
%&=ig\int d\mathbf{z}_{\perp} \mathrm{Tr}\left(U(\mathbf{z}_{\perp}) T^b \frac{\delta}{\delta U^{\dagger}(\mathbf{z}_{\perp})}\right) \left[\frac{1}{\partial D} \right]^{ba}(\mathbf{z}_{\perp} , \mathbf{x}_{\perp})\\
= -ig \int d^2\mathbf{z}_{\perp}  \left[\frac{1}{ D\partial} \right]^{ae}(\mathbf{x}_{\perp} , \mathbf{z}_{\perp})\mathcal{J}^e_R(\mathbf{z}_{\perp})\, 
%\end{split}
\end{equation}
with the right rotation operator
\begin{equation}
\mathcal{J}^b_R(\mathbf{z}_{\perp})  = -\mathrm{Tr}\left(U(\mathbf{z}_{\perp}) T^b \frac{\delta}{\delta U^{\dagger}(\mathbf{z}_{\perp})}\right)\, 
\end{equation}
This operator  when acting on $U$ multiplies it  on the right by $T^b$. 
Also recall that
\begin{equation}
\partial_i \frac{1}{\partial^2}(\mathbf{x}_{\perp}, \mathbf{y}_{\perp}) = \frac{1}{2\pi} \frac{(\mathbf{x}_{\perp}-\mathbf{y}_{\perp})_i}{(\mathbf{x}_{\perp}-\mathbf{y}_{\perp})^2}\,, \quad D_i \frac{1}{D^2}(\mathbf{x}_{\perp}, \mathbf{y}_{\perp}) = \frac{1}{2\pi}U^{\dagger}(\mathbf{x}_{\perp}) \frac{(\mathbf{x}_{\perp}-\mathbf{y}_{\perp})_i}{(\mathbf{x}_{\perp}-\mathbf{y}_{\perp})^2}U(\mathbf{y}_{\perp})\,. 
\end{equation}
Finally, eq. \eqref{eq:BL_BR1} becomes
\begin{equation}\label{qu}
Q_i^a[\mathbf{x}_{\perp};U]  = -\frac{g}{2\pi}\int d^2\mathbf{y}_{\perp} \frac{(\mathbf{x}_{\perp}-\mathbf{y}_{\perp})_i}{(\mathbf{x}_{\perp}-\mathbf{y}_{\perp})^2} \left[ U(\mathbf{y}_{\perp}) -U(\mathbf{x}_{\perp})\right]^{ab} \mathcal{J}_R^b(\mathbf{y}_{\perp})
\end{equation}
The standard JIMWLK kernel is reproduced as
\begin{equation}
\begin{split}\label{jimwlk}
H_{JIMWLK}\,=& -\frac{1}{2\pi} \int d^2\mathbf{z}_{\perp} Q_i^{a\dagger}[\mathbf{z}_{\perp}; U] Q_i^a[\mathbf{z}_{\perp}; U] \\
=& - \frac{\alpha_s}{2\pi^2} \int_{\mathbf{z}_{\perp}, \mathbf{y}_{\perp},\mathbf{x}_{\perp}}\frac{(\mathbf{z}_{\perp} - \mathbf{x}_{\perp})_i}{(\mathbf{z}_{\perp} - \mathbf{x}_{\perp})^2}\frac{(\mathbf{z}_{\perp} - \mathbf{y}_{\perp})_i}{(\mathbf{z}_{\perp} - \mathbf{y}_{\perp})^2}\mathcal{J}^e_{R}(\mathbf{y}_{\perp})\\
&\times [ 1+ U^{\dagger}(\mathbf{y}_{\perp}) U(\mathbf{x}_{\perp}) -  U^{\dagger}(\mathbf{y}_{\perp}) U(\mathbf{z}_{\perp}) - U^{\dagger}(\mathbf{z}_{\perp}) U(\mathbf{x}_{\perp})]^{ed} \mathcal{J}^d_R(\mathbf{x}_{\perp})\, .
\end{split}
\end{equation}
If \eqref{wj} is expanded  to leading order in $j$ rather than $\tau$, one obtains the so called KLWMIJ equation, which describes the scattering of a dilute projectile on a dense target. It is dual to the JIMWLK equation \eqref{jimwlk} and is obtained from it by the Dense-Dilute Duality transformation $U(\mathbf{x}_\perp)\leftrightarrow {\cal R}_p(\mathbf{x}_\perp)$ \cite{Kovner:2005en}.

This completes our recap of the JIMWLK equation. 

\section{The "CSS-JIMWLK" Hamiltonian}

Our goal now is to derive the equation that governs the evolution of another set of variables, i.e. those that involves gluons at fixed value of rapidity $\ln1/x$ even as the rapidity "resolution" is finer, or in other words the wave function has been evolved past the rapidity of these gluons to $\eta\gg \ln1/x$. The total evolution interval $\eta$ clearly provides the longitudinal resolution in the same sense as usually discussed in the context of longitudinal resolution of TMD, 
as the softest gluons present in a wave function evolved to $\eta$, have rapidity $\eta$.
We note that technically the "factorization scheme" inherently provided by the evolution discussed here is a little different than in the standard approach. However to leading order in $\alpha_s$ to which we are working here this does not matter.

In the context of JIMWLK equation, the evolution in $\eta$ physically represents evolution in the total energy of the process. In the context of the present work however one can think about it alternatively as an evolution in the longitudinal factorization scale, which for some observables does not have to be identical with the total energy.  It is useful to keep this distinction in mind in the following.
In both cases however, the evolution is generated by  boosting the hadronic wave function and probing its gluonic structure.
Formally it is not difficult to derive the equation we are after.% In the language of the wave function evolution this corresponds to the "Lindblad" terms which collects the virtual contribution and the contribution where the gluons emitted into the new rapidity interval do not directly contribute to the observable one is calculating.

The starting point of the derivation is again \eqref{deltarho} - the choice of observable does not affect the evolution of the wave function, or density matrix.   Now however  \eqref{ohat} is modified. We are not interested in an "JIMWLK-like" observable $\hat O$ that depends only  on the valence charge, and certainly not on the charge integrated over the whole range of rapidity all the way to $\eta$. Instead the observable $\hat {\cal O}$, whatever it is, stays the same throughout the evolution  and  its  expectation value changes  solely due to the evolution of the soft gluon wave function. In other words, if for the JIMWLK we integrated over soft gluons only partially, keeping track of their contribution to $j$ which modified the observable, now we should integrate out the soft gluons completely, in the Wilsonian sense, since the observable does not depend on these degrees of freedom. We note that this is a typical situation considered in the physics of open quantum systems, which generally (if the system is Markovian) leads to the Kossakowski-Lindblad equation  for time evolution of reduced density matrix \cite{Gorini:1975nb,Lindblad:1975ef}.

In practical terms  this means that the operator $\hat R$ in \eqref{ohat} should be set equal to unity. Retracing  our steps in the derivation of JIMWLK we see that this leads
to one "minor" modification in the evolution kernel in \eqref{wj}. Namely  $\tilde b_L$ should be replaced by $b_L$, since
 %the operator 
 $\mathcal{R}_p$ in the definition of $\tilde b_L$ originates precisely from the operator $\hat R$ in \eqref{ohat}. Note that we still have two "classical fields"  $b_L$ and $b_R$, since the index here differentiates between the fields in the wave function and the conjugate wave function that are ordered in a specific way relative to the operator $\hat {\cal O}$\cite{footnote}.

With this modification the new "master Hamiltonian" is:
\begin{equation}\label{eqpr}
H'_{RFT} =\frac{1}{4\pi} \Bigg[({b}_L^{a} - b_R^{a})^{\dagger}(N^{\dagger}_{\perp,R}N_{\perp,R} )^{ab}({b}_L^{b} -b_R^{b}) +h.c.\Bigg]
%H'_{RFT} =\frac{1}{4\pi} \Bigg[({b}_L^{\alpha} - b_R^{\alpha})^{\dagger}(N^{\dagger}_{\perp,R}N_{\perp,R} )_{\alpha\beta}({b}_L^{\beta} -b_R^{\beta}) +h.c.\Bigg]
\end{equation}
As before, this Hamiltonian is applicable for either dilute target or dilute projectile. We first explore the limit of dense projectile and dilute target, i.e. expand the Hamiltonian \eqref{eqpr}  to second order in $\tau$.
\begin{equation}
\begin{split}
&b_L^a - b_R^a  = b_i^a[j_L]  - b_i^a[j_R] \\
\simeq & \left(b_i^a[j] + \frac{\delta b_i^a}{\delta j^e}(j_L^e -j^e)\right) - \left(b_i^a[j] +\frac{\delta b_i^a}{\delta j^e} (j_R^e -j^e)\right)\\
\simeq& \frac{\delta b_i^a}{\delta j^e}g j^cT^d_{ce}\frac{\delta}{\delta j^d}=\int d^2\mathbf{z}_{\perp}\frac{\delta b_i^a(\mathbf{x}_{\perp})}{\delta j^e(\mathbf{z}_{\perp})} gj^c(\mathbf{z}_{\perp})T^d_{ce}\frac{\delta}{\delta j^d(\mathbf{x}_{\perp})}\\
=&i\left[D_i - D_i\frac{1}{\partial D} D\partial \right]^{ad}(\mathbf{x}_\perp,\mathbf{z}_\perp)\, \frac{\delta}{\delta j^d(\mathbf{z}_{\perp})}
\end{split}
\end{equation}
A little bit more algebra and
\begin{eqnarray}\label{nb}
[N_\perp(b_L-b_R)]^a\simeq\partial_i\frac{1}{\partial^2}(D\partial-\partial D)^{ab} (\mathbf{x}_\perp,\mathbf{z}_\perp)\frac{\delta}{\delta j^b(\mathbf{z}_{\perp})}=ig\partial_i\frac{1}{\partial^2}(\mathbf{x}_\perp,\mathbf{z}_\perp)\left[j(\mathbf{z}_\perp)T^a\frac{\delta}{\delta j(\mathbf{z}_\perp)}\right]\nonumber \\
%=ig\partial_i\frac{1}{\partial^2}(\mathbf{x},\mathbf{z})\mathcal{I}^V_a(\mathbf{z})\nonumber
\end{eqnarray}
Thus the CSS -JIMWLK Hamiltonian for a dense projectile can be conveniently written as
\begin{eqnarray}\label{css1}
H_{CSS-JIMWLK}=%\frac{g^2}{2\pi}\int_{\mathbf{x},\mathbf{y}}\mathcal{I}^V(\mathbf{x})\frac{1}{\partial^2}(\mathbf{x},\mathbf{y})\mathcal{I}^V(\mathbf{y})\\
\frac{g^2}{2\pi}\int_{\mathbf{x_\perp,y_\perp}}\left[j(\mathbf{x}_\perp)T^a\frac{\delta}{\delta j(\mathbf{x}_\perp)}\right]\left[\frac{1}{\partial^2}\right](\mathbf{x_\perp,y_\perp})\left[j(\mathbf{y_\perp})T^a\frac{\delta}{\delta j(\mathbf{y}_\perp)}\right]
\end{eqnarray}

We can now perform the same exercise for a dilute projectile. This time we expand \eqref{eqpr} to lowest order in $j$ but keep all orders in $\tau$. Interestingly, the result we obtain is identical to \eqref{css1}. We thus discover that the CSS-JIMWLK Hamiltonian we have derived is universal and is applicable equally to dilute and dense hadronic system.

Another way of writing $H_{CSS-JIMWLK}$ is in terms of the 
 dual charge rotation operators $\mathcal{I}_{R(L)}$. These are analogous to $\mathcal{J}_{R(L)}$ but act as rotation operators on the matrix ${\cal R}_p$ rather than $U$:
\begin{equation}
\mathcal{I}^b_R(\mathbf{z}_{\perp})  = -\mathrm{Tr}\left({\cal R}_p(\mathbf{z}_{\perp}) T^b \frac{\delta}{\delta {\cal R}_p^{\dagger}(\mathbf{z}_{\perp})}\right)\, ; \ \ \ \ \ \ \ \mathcal{I}^a_L(\mathbf{z}_\perp)={\cal R}_p^{ab}(\mathbf{z}_\perp)\mathcal{I}^b_R(\mathbf{z}_\perp);\ \ \ \ \ \mathcal{I}^V\equiv \mathcal{I}_L-\mathcal{I}_R
\end{equation}
\begin{eqnarray}\label{css2}
H_{CSS-JIMWLK}=\frac{g^2}{2\pi}\int_{\mathbf{x}_\perp,\mathbf{y}_\perp}\mathcal{I}^V(\mathbf{x}_\perp)\frac{1}{\partial^2}(\mathbf{x}_\perp,\mathbf{y}_\perp)\mathcal{I}^V(\mathbf{y}_\perp)
%\frac{g^2}{2\pi}\int_{x,y}\left[j(\mathbf{x})T^a\frac{\delta}{\delta j(\mathbf{x})}\right]\left[\frac{1}{\partial^2}\right](\mathbf{x,y})\left[j(\mathbf{y})T^a\frac{\delta}{\delta j(\mathbf{y})}\right]
\end{eqnarray}

%At any rate \eqref{css1} has a very simple form with a rather intuitive explanation.
 %A dense target contains many gluons, but only two of them scatter on a dilute projectile. Eq.\eqref{css1} simply says that the evolution consists of emission of another gluon from either one of the two that scatter, but without this additional gluon actually scattering. The fact that the additional gluon does not scatter is quite natural, since as per our setup it does not contribute directly to the observable. It is a little surprising though, that there are no nonlinear effects in the emission of this gluon and it is thus insensitive to the existence of spectator gluons in the dense wave function, but only to the small number of gluons that participate in the scattering.
 
 Eqs. \eqref{css1} and \eqref{css2} have a very simple form and its physics is easy to understand.  The operator $jT^a\frac{\delta}{\delta j}$ is  an emission vertex of a gluon from the charge density $j$.  Recall that the charge density $j$ is integrated over rapidity up to the longitudinal resolution scale $ \eta$, thus  $jT^a\frac{\delta}{\delta j}$ is a gluon emission vertex from {\it any} gluon  with rapidity $\eta_g<\eta$ present in the wave function, since all such gluons contribute to $j$. Thus the action of $H_{CSS-JIMWLK}$  amounts to emission  of a soft gluon in the wave function and its reabsorption (either in the wave function or its conjugate) from any of the "valence" gluons in the wave function. 
 
 In the Wigner-Weyl representation
 \begin{equation}
 \langle \eta|\hat {\cal O}|\eta \rangle\,=\,\int dj\, W_\eta[j] \,\tilde {\cal O}(j)\,\equiv\,\langle\, \tilde  {\cal O}\, \rangle_\eta\
 \end{equation}
 Algebraically, the action of $jT^a\frac{\delta}{\delta j}$ on 
 the operator $\tilde {\cal O}(j)$ is nothing but a color rotation of any of the valence gluon creation and annihilation operators.
 The evolution equation, generated by $H_{CSS-JIMWLK}$ is
\begin{equation}
\frac{d}{d\eta}\langle \eta|\hat {\cal O}|\eta \rangle=
%-\langle \eta| H_{CSS} \hat {\cal O}|\eta \rangle\,=\,
-\langle  \,H_{CSS-JIMWLK}\,\tilde {\cal O}\,\rangle_\eta
\end{equation}
The action of $H_{CSS-JIMWLK}$ on 
%any operator 
$\tilde {\cal O}$ follows from the action of $jT^a\frac{\delta}{\delta j}$. As we have discussed above, $j(\mathbf{x}_\perp)T^a\frac{\delta}{\delta j(\mathbf{x}_\perp)}$ is the Wigner-Weyl representation of the color rotation operator \cite{Agarwal:1971wc}, therefore
\begin{equation}\label{jo}
\langle\, j(\mathbf{x}_\perp) T^a\frac{\delta}{\delta j(\mathbf{x}_\perp)}\tilde {\cal O}\,\rangle_\eta \equiv \langle \eta|[\hat j^a(\mathbf{x}_\perp),
\hat {\cal O}]|\eta\rangle\,.
\end{equation}
To be mathematically precise we would have to find a Wigner-Weyl representation $\tilde {\cal O}$ of $\hat {\cal O}$, but we are not going to do it here since the exact form depends on the operator $\hat {\cal O}$ while \eqref{jo} holds for all operators of the type we are discussing.

The evolution equation then is
\begin{equation}\label{opev}
\frac{d}{d\eta}\langle\eta\,| \hat {\cal O}|\,\eta\rangle=-\frac{g^2}{2\pi}\langle \eta|\,\int d^2\mathbf{u}_\perp d^2\mathbf{v}_\perp\frac{1}{\partial^2}(\mathbf {u}_\perp,\mathbf{v}_\perp)[\hat j^a(\mathbf{u}_\perp),[\hat j^a(\mathbf{v}_\perp),\hat {\cal O}]]\,|\eta\rangle
\end{equation}
This form of the evolution is the most straightforward to use for generic operators $\hat {\cal O}$. This equation is of the type of Kossakowski-Lindblad equation \cite{Gorini:1975nb,Lindblad:1975ef} ubiquitous in discussions of dynamics of open quantum systems, even though in our case the dynamics is in rapidity rather than in time.

One might wonder why did we go into trouble of working with the Wigner-Weyl representation for the evolution in the first place, if we are going to use the operator representation \eqref{opev} at the end of the day. In fact, in the dilute case,  \eqref{opev} can be easily derived directly using the wave function \eqref{wf} with \eqref{eq:coherent_operator}. However for dense case this derivation albeit possible, is considerably more cumbersome. The advantage of our approach here is that it yields equally easily $H_{CSS-JIMWLK}$ in both dilute and dense cases.

 \section{The evolution of the gluon TMD  and the CSS equation}
 
Let us see in  more detail how  $H_{CSS-JIMWLK}$ evolves  some basic observables. As the simplest observable let us  consider the unpolarized gluon TMD. 
The unpolarized gluon TMD in the most naive way is defined as the correlator of the gluon fields at transverse momentum $\mathbf{k}$. 
In addition we have to be clear about the longitudinal momentum fraction and the longitudinal resolution dependence of the TMD.  We are interested in $T(\mathbf{k},x,\zeta)$, where $\mathbf{k}$ denotes the transverse momentum of the TMD, $x=k^+/P^+$ denotes the longitudinal momentum fraction and $\zeta$ denotes the longitudinal resolution parameter. As noted earlier we use a dimensionless longitudinal resolution parameter rather than a dimensional one usually introduced in CSS. The relation between the two is logarithmic, and to leading order in $\alpha_s$ the two are equivalent. We stress that here we define TMD as the number of gluons at a given transverse and longitudinal momentum at a fixed resolution. In any calculation of physical process the most convenient choice of the resolution scale is dictated by the kinematics of the process, and we are not considering this question here.

We do not introduce in this discussion the transverse resolution parameter $\mu^2$, since as we shall see in the standard CGC approach this parameter is naturally associated with the UV cutoff.

\subsection{The CSS equation}

Given our previous discussion, it is clear how $x$ and $\zeta$ are accommodated in the high energy evolution framework. From the point of view of the high energy evolution, the hadronic wave function evolved to a particular value of rapidity $\eta$ contains only gluons with longitudinal momentum $k^+\le P^+e^{-\eta}$. Thus the parameter of the evolution $\eta$ naturally provides the longitudinal resolution in the same sense as longitudinal resolution in the perturbative TMD physics. Therefore the longitudinal resolution of a TMD calculated in the hadronic wave function evolved to rapidity $\eta$, is in fact $\zeta=\eta$. 
The operator definition of the TMD is therefore
\begin{equation}\label{tmd}
T(\mathbf{k},x,\zeta)=\langle \zeta | \,\hat T(\mathbf{k},x)\,|\zeta\rangle\,;~~~~~~~~~\hat T(\mathbf{k},x)=\int d^2\mathbf{x}_\perp d^2\mathbf{y}_\perp\, e^{i\mathbf{k\cdot(x_\perp-y_\perp)}}\  a^{\dagger a}_i(x,\mathbf{x}_\perp)\,a^a_i(x,\mathbf{y}_\perp)
%\int d^2\mathbf{x}_\perp d^2\mathbf{y}_\perp\, e^{i\mathbf{k\cdot(x_\perp-y_\perp)}}\ \langle \zeta | a^{\dagger a}_i(\mathbf{x}_\perp,x)\,a^a_i(\mathbf{y}_\perp,x)|\zeta\rangle
\end{equation}
where $|\zeta\rangle$ is the hadronic state evolved to rapidity $\eta=\zeta$. 
%For simplicity, we define the TMD operator 
%$\hat T(\mathbf{k},x)=\int d^2\mathbf{x}_\perp d^2\mathbf{y}_\perp\, e^{i\mathbf{k\cdot(x_\perp-y_\perp)}}\  a^{\dagger a}_i(\mathbf{x}_\perp,x)\,a^a_i(\mathbf{y}_\perp,x)$

It is obvious that  gluon TMD at longitudinal momentum fraction $x$  vanishes as long as $\eta\le \ln 1/x$. When the wave function is evolved up to  $\eta=\ln 1/x$, the gluon TMD is generated for the first time by splittings of the valence gluons. At this point in the evolution, the TMD is given  by the correlator of the Weizsacker-Williams field \cite{feng}
\begin{equation}\label{tin}
\begin{split}
T(\mathbf{k},x,\eta=\ln 1/x)&=\,\int d^2\mathbf{x}_\perp d^2\mathbf{y}_\perp\,e^{i\mathbf{k\cdot(x_\perp-y_\perp)}}\langle \eta |\hat b^a_i(\mathbf{x}_\perp)\hat b^a_i(\mathbf{y}_\perp)|\eta \rangle \\ \nonumber 
&=\,\int d^2\mathbf{x}_\perp d^2\mathbf{y}_\perp\,e^{i\mathbf{k\cdot(x_\perp-y_\perp)}}\, 
\int dj\,W_\eta[j]\,b^a_i(\mathbf{x}_\perp)\, b^a_i(\mathbf{y}_\perp)
\end{split}
\end{equation} 
where
 %the averaging denotes expectation value in the hadronic state evolved to rapidity $\eta$, and 
 the gluon field $ b$ is determined through \eqref{bj} by the color charge density $ j$ integrated over 
 rapidity up to $\eta=\ln 1/x$\cite{footnote2}. 

Now consider further evolution in $\eta$. Naively, one could think that one has to evolve the product of Weizsacker - Williams fields (or equivalently $W_\eta$) with $H_{JIMWLK}$, but this is incorrect. 
The $\eta$  evolution increases the longitudinal resolution in the wave function, but does not change the longitudinal momentum fraction $x$ in the TMD. 
On the other hand the evolution with $H_{JIMWLK}$ increases the resolution but also decreases the value of the momentum fraction $x$. 
The TMD at fixed $x$ and $\zeta>\ln 1/x$  has to be calculated by evolving with the Hamiltonian $H_{CSS-JIMWLK}$ \eqref{css1} in $\eta$ from $\eta=\ln 1/x$ to $\eta=\zeta$ with  the initial condition \eqref{tin}. 

%To summarize, the way to calculate $T(\mathbf{k},x,\zeta)$ in the high energy evolution approach one needs first to evolve the charge density with JIMWLK Hamiltonian to rapidity $\eta_0=\ln 1/x$, and subsequently evolve the TMD defined in \eqref{tin} with the CSS Hamiltonian \eqref{css1} to rapidity $\eta=\zeta$. To be clear, the second step in this procedure generates 
%TMD as a functional of the color charges integrated to $\eta_0$, i.e. $j_{\eta_0}$. The functional form of TMD in terms of $j_{\eta_0}$ depends on the length of this second step, i.e. on $\zeta-\eta_0$. Once this functional form is obtained, the averaging over $j_{\eta_0}$ has to be performed with the weight function $W_{\eta_0}[j]$ determined by the JIMWLK evolution to rapidity $\eta_0$.

Acting with the Hamiltonian \eqref{css1} on the operator $\hat T(\mathbf{k},x)$ %$a^{\dagger a}_i(\mathbf{x},x)a^a_i(\mathbf{y},x)$   
leads to a simple homogeneous equation for the TMD.
\begin{equation}
\begin{split}
\frac{\partial T(\mathbf{k},x,\eta)}{\partial \eta}=-\frac{g^2}{2\pi}\int d^2\mathbf{u}_\perp d^2\mathbf{v}_\perp\frac{1}{\partial^2}(\mathbf {u}_\perp,\mathbf{v}_\perp)\,\langle\eta|[\hat j^a(\mathbf{u}_\perp),[\hat j^a(\mathbf{v}_\perp),\hat T(\mathbf{k},x)]] |\eta\rangle 
\end{split}
\end{equation}
To evaluate the right-hand side, we use 
\begin{equation}
\begin{split}
[\hat j^a(\mathbf {v}_\perp), a^{\dagger\,b }_{i}(x,\mathbf {x}_\perp)]&=\Big[a_i^{\dagger}(x, \mathbf {v}_\perp)T^a\Big]^b\delta^{(2)}(\mathbf {x}_\perp-\mathbf {v}_\perp)
\end{split}
\end{equation}
and similarly for the commutator with $a$.
This of course follows from the canonical commutation relation
%note how charge operator acts on the $a(p^+,\mathbf {x}_\perp)$ or $a^\dagger(p^+, \mathbf {x}_\perp)$. We use the following convention for the commutator,
\begin{equation}
\begin{split}
 [a_i^a(p^+,\mathbf {x}_\perp), a_j^{\dagger b}(k^+,\mathbf {y}_\perp)]=(2\pi)\delta(p^+-k^+)\delta_{ij}\delta_{ab}\delta^{(2)}(\mathbf {x}_\perp-\mathbf {y}_\perp)
\end{split}
\end{equation}
Two consecutive color rotations yield 
 %at the same location,
\begin{equation}
\begin{split}
[\hat j^a(\mathbf {u}_\perp),[\hat j^a(\mathbf {v}_\perp), a_{i}^{\dagger\,b}(x,\mathbf {x}_\perp)]]&=N_ca_{i}^{\dagger\,b}(x, \mathbf {u}_\perp)\delta^{(2)}(\mathbf {x}_\perp-\mathbf {v}_\perp)
\end{split}
\end{equation}
The final result is
\begin{eqnarray}\label{css0}
&&\frac{\partial T(\mathbf{k},x,\eta)}{\partial \eta}
=-\frac{2g^2N_c}{2\pi}\int d^2\mathbf{u}_\perp d^2\mathbf{v}_\perp \int d^2\mathbf{x}_\perp d^2\mathbf{y}_\perp\ 
e^{i\mathbf{k\cdot(x_\perp-y_\perp)}}\frac{1}{\partial^2}(\mathbf {u}_\perp,\mathbf{v}_\perp) \nonumber \\
&&\times\Big[\delta^{(2)}(\mathbf {u}_\perp-\mathbf {v}_\perp)\delta^{(2)}(\mathbf {v}_\perp-\mathbf {x}_\perp)-\delta^{(2)}(\mathbf {u}_\perp-\mathbf {x}_\perp)\delta^{(2)}(\mathbf {v}_\perp-\mathbf {y}_\perp)     \Big]\langle \eta | a^{\dagger a}_i(x,\mathbf{x}_\perp)a^a_i(x,\mathbf{y}_\perp)|\eta\rangle \nonumber \\
&&= -\frac{2\alpha_sN_c}{2\pi} \int d^2\mathbf{x}_\perp d^2\mathbf{y}_\perp\ e^{i\mathbf{k\cdot(x_\perp-y_\perp)}} \,\langle \eta | a^{\dagger a}_i(x,\mathbf{x}_\perp)a^a_i(x,\mathbf{y}_\perp)|\eta\rangle
\nonumber \\
& &\times \Big[\int d^2\mathbf {u}_\perp \ln \frac{1}{(\mathbf {u}_\perp-\mathbf{x}_\perp)^2 \lambda^2}\delta^{(2)}(\mathbf {u}_\perp-\mathbf {x}_\perp)-\ln \frac{1}{(\mathbf {x}_\perp-\mathbf{y}_\perp)^2 \lambda^2}\Big]
\end{eqnarray}
where we have used,
\begin{equation}
\frac{1}{\partial^2}(\mathbf {u}_\perp,\mathbf{v}_\perp)=\frac{1}{4\pi} \ln \frac{1}{(\mathbf {u}_\perp-\mathbf{v}_\perp)^2 \lambda^2}
\end{equation}
with $\lambda$ - an IR cut-off. The result does not depend on the choice of $\lambda$ due to the cancellation between real and virtual terms. The integration over $\mathbf {u}_\perp$ in the virtual contribution (second term in \eqref{css0} simply introduces a UV cut-off $\Lambda$.
\begin{equation}
\int d^2\mathbf {u}_\perp \ln \frac{1}{(\mathbf {u}_\perp-\mathbf{x}_\perp)^2 \lambda^2}\delta^{(2)}(\mathbf {u}_\perp-\mathbf {x}_\perp)=\ln \frac{\Lambda^2}{\lambda^2}\,.
\end{equation}
Since the TMD is a decreasing function of the transverse momentum and logarithm is a slowly varying function, the real term can be approximated as
\begin{equation}\label{appr}
\begin{split}
&\hspace{-2cm}\int d^2\mathbf{x}_\perp d^2\mathbf{y}_\perp\ e^{i\mathbf{k\cdot(x_\perp-y_\perp)}} \ln \frac{1}{(\mathbf {x}_\perp-\mathbf{y}_\perp)^2 \lambda^2} \langle \eta | a^{\dagger a}_i(x,\mathbf{x}_\perp)a^a_i(x,\mathbf{y}_\perp)|\eta \rangle \\
&\approx  \ln \frac{\mathbf {k}^2}{ \lambda^2}
\int d^2\mathbf{x}_\perp d^2\mathbf{y}_\perp\ e^{i\mathbf{k\cdot(x_\perp-y_\perp)}} \langle \eta | a^{\dagger a}_i(x,\mathbf{x}_\perp)a^a_i(x,\mathbf{y}_\perp)|\eta\rangle
\end{split}
\end{equation}
In summary, \eqref{css0}  reduced to the simple momentum space form of the CSS equation \cite{CSS},
\begin{equation}\label{tmdcss}
\frac{\partial T(\mathbf{k},x,\eta)}{\partial \eta}=\frac{-2\alpha_s N_c}{2\pi}\ln\frac{\Lambda^2}{\mathbf{k}^2}\ T(\mathbf{k},x,\eta)
\end{equation}
with the transverse resolution scale $\mu^2=\Lambda^2$.
%where $\Lambda^2$ is the UV cutoff introduced to make the transverse momentum integral finite. 
%Eq. (\ref{tmdcss}) is obviously an approximate one. 

Without the approximation \eqref{appr},  Eq. (\ref{css0}), when Fourier transformed into coordinate space yields the standard perturbative form of the CSS equation in impact parameter space  as is commonly used in the TMD literature \cite{TMDbook}. 

The solution for this simple homogeneous equation is of course given by the doubly logarithmic Sudakov suppression factor
\begin{equation}\label{sol}
T(\mathbf{k};x,\zeta)=e^{-\frac{\alpha_s N_c}{\pi}\ln\frac{\Lambda^2}{\mathbf{k}^2}[\zeta-\ln 1/x]}T(\mathbf{k};x,\ln 1/x)
\end{equation}
with the initial condition \eqref{tin}.

\subsection{The transverse resolution}

Eq.\eqref{tmdcss} is the CSS equation for the evolution of the gluon TMD in the longitudinal resolution parameter with the transverse resolution scale equal to the UV cutoff. It is quite natural  that the UV cutoff arises as  the transverse resolution scale in the present framework. The reason is that in the JIMWLK approach, where the evolution is with respect to the logarithm of the longitudinal momentum, all transverse momentum modes, up to the UV cutoff populate the wave function at a fixed $\eta$. Therefore the transverse resolution for TMD defined in \eqref{tmd} is $\Lambda$. 

One could try to rectify this by introducing the resolution scale directly in the definition of the TMD. This seems to make perfect sense, as at the end of the day it is the physical process that determines which observable is being probed in the wave function. 
However this fails to modify the resolution scale in the evolution equation. Let us try to do it nevertheless to understand the problem.

Let us modify the definition of the TMD by hand to introduce a transverse resolution scale. To do that we introduce smeared creation and annihilation operators
\begin{equation}
a^{a\mu}_i(x,\mathbf{x}_\perp)\equiv\int_{{z}_\perp} a_i^a(x,\mathbf{x_\perp-z_\perp}) f^{ \mu}(\mathbf{z}_\perp)
\end{equation}
where the smearing function $f$ has the properties
\begin{equation}
\int d^2\mathbf{z}_\perp f^\mu(\mathbf{z}_\perp)=1; \ \ {\rm and} \ \ f^\mu(\mathbf{z}_\perp)\rightarrow 0; \ \ \ \mathbf{z}_\perp^2>1/\mu^2
\end{equation}
The exact shape of the function $f$ does not matter and amounts to the choice of the "transverse factorization scheme". In particular a simple Gaussian form $f(\mathbf{z}_\perp)=\frac{\mu^2}{\pi}e^{-\mu^2\mathbf{z}_\perp^2}$ is a convenient choice.
We further define the transverse momentum resolution dependent TMD 
\begin{equation}\label{tmdmu}
T(\mathbf{k},\mu^2;x,\zeta)=\int d^2\mathbf{x}_\perp d^2\mathbf{y}_\perp \ e^{i\mathbf{k\cdot(x_\perp-y_\perp)}}\ \langle \zeta | a^{\dagger a \mu}_i(x,\mathbf{x}_\perp)a^{a\mu}_i(x,\mathbf{y}_\perp)|\zeta\rangle
\end{equation}
This definition is eminently reasonable. Now let us consider its evolution  with $H_{CSS-JIMWLK}$.
The UV divergence in the calculation above arose from the term where both charge densities $j$ in $H_{CSS-JIMWLK}$ hit the same creation or annihilation operators, setting the coordinates of the two charge densities to be equal, thereby bringing the UV cutoff into the game. With the smeared gluon field the problem unfortunately is not going away. When acting with $H_{CSS-JIMWLK}$ on a single smeared field we still get the contribution only when the coordinates of $\hat j(\mathbf{u}_\perp)$  and $\hat j(\mathbf{v}_\perp)$ are equal, and thus this contribution is proportional to $\lim_{|\mathbf{u}_\perp-\mathbf{v}_\perp|\rightarrow 0}\ln |\mathbf{u}_\perp-\mathbf{v}_\perp|$.  This yields the same UV divergence as before which is not regulated by $\mu$. 

Physically it is quite clear what is going on. The evolution in $k^+$ allows eikonal emissions with arbitrary transverse momenta of "soft" gluons. Thus in one step of the evolution the transverse phase space for emissions is infinite and the probability of a given gluon at longitudinal momentum $x$ to decay diverges. Kinematic restriction in the definition \eqref{tmdmu} does not change this fact.
The only way to regulate this UV divergence would be to smear the charge density $j$ that enters $H_{CSS-JIMWLK}$. However this is not in our power since $H_{CSS-JIMWLK}$ is determined unambiguously by the evolution of the wave function. On the other hand, were we to change the evolution parameter from $k^+$ to $k^-$ the situation would be quite different. This would limit the transverse space for emissions and indeed regulate the UV divergence at a transverse scale determined by $x$ and the resolution $\zeta$ \cite{inprep}. The evolution parameter would then simultaneously determine the longitudinal and transverse resolution as was observed in \cite{balitsky-tarasov}.

%\begin{equation}\label{tmdcss}
%\frac{\partial T(\mathbf{k},\mu^2;x,\zeta)}{\partial \zeta}=-\frac{\alpha_s N_c}{2\pi}\ln\frac{\mu^2}{\mathbf{k}^2}\ T(\mathbf{k},\mu^2;x,\zeta)
%\end{equation}

To conclude this section we make the following comment. Since at $\eta=\ln 1/x$ the TMD is given by \eqref{tin}, one might be naturally tempted to simply evolve this expression by acting on it directly with $H_{CSS-JIMWLK}$ \eqref{css1},
\begin{equation}
{d\tilde O\over d\eta}\,=\,H_{CSS-JIMWLK}\,\tilde O\,;~~~~~~~~~~ \tilde O\,\sim \, b^a b^a
\end{equation}
This is however incorrect. Such a procedure would amount to CSS-evolving the charge density integrated up to $\ln 1/x$, let us call it $j_x$, to rapidity higher than $\ln 1/x$  without adding to $j_x$ the charge of softer gluons, and then calculating the classical field $b_x$ determined by the evolved $j_x$ via classical Yang-Mills equations. This is not the correct evolution of TMD given in Eq. (\ref{css0}). Such an evolution would take into account correctly emission of the softest gluons from higher $x$  gluons that comprise $j_x$ and thus decrease in the number of such gluons \cite{footnote3}. These however are not the gluons with the longitudinal momentum fraction $x$, but rather those with the momentum fraction greater than $x$! The gluons at $x$ themselves also split, and it is this splitting that is first and foremost responsible for the decrease in their number via the Sudakov form factor. This splitting is not accounted for by simply CSS - evolving  $j_x$. To reiterate, the correlator of classical fields equals to the gluon TMD {\bf only} at rapidity $\eta=\ln 1/x$. At $\eta<\ln 1/x$, the TMD vanishes, while at $\eta>\ln 1/x$ it is not given by the correlator of classical fields created by the charge density $j_x$, but is suppressed by the Sudakov form factor (see (\ref{sol})).

 \subsection{The  quark TMD}
 
 The previous discussion can be extended easily to other similar quantum observables. As another example consider the quark TMD. Although in the high energy framework the quark TMD itself arises only at higher order in $\alpha_s$ relative to the gluon TMD, its evolution is a leading order effect.
 The unpolarized quark TMD is defined similarly to the unpolarized gluon TMD
 \begin{equation}\label{tmdq}
T_q(\mathbf{k},x,\zeta)=\langle \zeta | \,\hat T_q(\mathbf{k},x)\,|\zeta\rangle\,;~~~~~~\hat T_q(\mathbf{k},x)=\int d^2\mathbf{x}_\perp d^2\mathbf{y}_\perp\, e^{i\mathbf{k\cdot(x_\perp-y_\perp)}}\   d^{\dagger\alpha}_\sigma(x,\mathbf{x}_\perp)\,
d_\sigma^\alpha(x,\mathbf{y}_\perp)
%\int d^2\mathbf{x}_\perp d^2\mathbf{y}_\perp\, e^{i\mathbf{k\cdot(x_\perp-y_\perp)}}\ \langle \zeta | a^{\dagger a}_i(\mathbf{x}_\perp,x)\,a^a_i(\mathbf{y}_\perp,x)|\zeta\rangle
\end{equation}
with $d^\dagger$ and $d$ - quark creation/annihilation operators ($\alpha$ is a fundamental color index and $\sigma$ is a spinor index).

%It is straightforward to generalize our derivation of CSS to include quarks. 
In the presence of quarks $H_{CSS-JIMWLK}$ is a functional  of valence charge density, that also includes quarks.
%For the gluonic sector the latter is defined in (\ref{wf}). The total valence charge density additionally includes contribution from the quarks:
\begin{equation}
\hat j^a\,=\,\hat j_g^a\,+\,\hat j^a_{q}\,;~~~~~~~~
\hat j^a_q(\mathbf{x}_{\perp})\,=\,\int_{\Lambda^+}^\infty {dk^+\over 2\pi}\, d^\dagger(k^+,\mathbf{x}_{\perp})\,t^a\,d(k^+,\mathbf{x}_{\perp})
\end{equation}
Here $j_g$ is defined by \eqref{wf} and $t^a$ is a generator in fundamental representation of the color group.

The charge density $j_q^a$ acts on the quark operators as a color rotation:
 \begin{equation}
[\hat j^a_q(\mathbf {v}_\perp), d_\sigma^{\alpha}(x,\mathbf {x}_\perp)]\,=\, d_\sigma^{\beta}(x, \mathbf {v}_\perp)\,t^a_{\beta\alpha}\,\delta^{(2)}(\mathbf {x}_\perp-\mathbf {v}_\perp)
\end{equation}
and similarly on $d^\dagger$.  Hence, the CSS-JIMWLK evolution for quark TMD, up to the change $N_c \rightarrow C_F=(N_c^2-1)/2N_c$, 
is identical to the  gluon one:
\begin{equation}\label{tmdcssq}
\frac{\partial T_q(\mathbf{k},x,\eta)}{\partial \eta}\,\simeq\,-\frac{\alpha_s C_F}{\pi}\ln\frac{\Lambda^2}{\mathbf{k}^2}\ T_q(\mathbf{k},x,\eta)
\end{equation}
which is again the longitudinal CSS evolution equation.

\section{Discussion}
In this note we have discussed the evolution with energy of a set of observables that are defined at a fixed, albeit small value of the longitudinal momentum fraction $x$. These observables are distinct from the ones described by the JIMWLK equation. The latter governs the evolution of observables that formally are functions of the color charge integrated over the whole range of rapidity of gluons present in the wave function. These observables are  dominated by the softest gluons present in the wave function at high energy.
We have derived the Hamiltonian- $H_{CSS-JIMWLK}$ that is appropriate for evolving the "non-JIMWLK" observables with energy.

The Hamiltonian $H_{CSS-JIMWLK}$ was obtained in the dilute and dense projectile limits. Interestingly, the form of the Hamiltonian in both limits is identical, and we believe on physical grounds that this form should be universal beyond these two limits. This may sound surprising at  first sight. In both limits the Hamiltonian has a simple form of the leading order emission probability of a single soft gluon. In the dilute case this is very natural, since this is the essence of the CSS physics - the gluon TMD depletes due to perturbative emissions of gluons into the new phase space opened up by evolution. In the dense case the basic physics is the same, but one could have expected  the emission probability to be affected by nonlinearities inherent in the dense wave function. However the fact that comes into play here, is that the wave function is "dense" only due to high density of the softest gluons - the gluons with the smallest possible longitudinal momentum fraction. Here  on the other hand we are interested in the emission off higher $x$ gluons, and those are not plentiful even in the highly evolved wave function. Although the soft gluon emitted from the TMD can interact with other soft gluons, this interaction does not affect the change of the state of the primary emitter. Thus the nonlinear corrections are irrelevant for the evolution of the TMD, and TMD-like observables. We believe this is the reason why $H_{CSS-JIMWLK}$ is unaffected by the finite gluon density effects included in JIMWLK.

Interestingly, the evolution equation generated by $H_{CSS-JIMWLK}$ is of the Kossakowski-Lindblad type \cite{Gorini:1975nb,Lindblad:1975ef}. As explained in \cite{Armesto:2019mna,ming}, the JIMWLK Hamiltonian also corresponds to the Kossakowski-Lindblad evolution of the reduced density matrix, albeit the form of the evolution is different. In the case of JIMWLK this is not obvious since one does not integrate the soft gluons from the density matrix fully, but keeps their contribution to the color charge. In the CSS case the situation is much cleaner - one fully integrates over the soft gluons in the evolution and thus the Kossakowski-Lindblad form of the evolution has exactly the same origin as in open quantum systems.

We have shown that when applied to gluon TMD, the evolution generated by $H_{CSS-JIMWLK}$ yields  the CSS evolution equation. To be more precise, we recover one of the two CSS equations - the evolution with respect to the longitudinal resolution scale. This  is of course the best we can hope for given that we only ever evolve with respect to one parameter. The dynamics of the transverse evolution is not formally included in the evolution of the wave function. 
The major drawback of the high energy evolution paradigm that gives rise to the standard JIMWLK equation is that the transverse resolution scale  is necessarily given by the UV cutoff. The reason is that the high energy evolution as formulated in the original BFKL and JIMWLK approach is the evolution in the longitudinal momentum. The transverse phase space is unrestricted and emissions of gluons with arbitrarily large transverse momenta are allowed. Therefore having finite evolution interval introduces a finite resolution in longitudinal momentum, but without restricting transverse momenta. 

This is not a problem fundamentally, but has to pose a significant inconvenience in practical terms. Any physical process  resolves the TMD at a finite transverse resolution scale, usually determined by the hard physical scale of the process. This means that using TMD defined with this particular resolution scale in TMD factorization formulae simplifies calculations significantly. It  makes the hard part of the process simple and perturbative with all relevant resummations of large logarithms contained in the calculation of TMD. One could in principle instead use TMD at a very high scale, i.e. the UV cutoff. But in this case the hard part will also contain this high scale, therefore it will contain large logarithms that will have to be resummed.  There should be large cancellations between the TMD and the hard part in the final result. This clearly is far from the optimal way to organize any calculation. Additionally we note that the leading order JIMWLK evolution contains only the double logarithmic part of the DGLAP splittings. 
In order to fully recover the DGLAP evolution, and therefore also the single Sudakov logarithm in CSS, one needs to resum higher order corrections in the JIMWLK Hamiltonian \cite{dglapbfkl}.  
Both of these problems are rectified if one considers the evolution in $k^-$ rather than $k^+$ \cite{balitsky-tarasov,inprep}.

%Thus we do expect to be able to recover the doubly logarithmic part of the second CSS equation, or alternatively the Sudakov double logarithm within our present framework.
%Referring back to \eqref{tmdcss}, and requiring that the TMD is indeed a well defined function of its variables we must have
%\begin{equation}
%\frac{\partial^2}{\partial\zeta\partial (\ln \mu^2)} T(\mathbf{k},\mu^2;x,\zeta)=\frac{\partial^2}{\partial \ln (\mu^2)\partial\zeta} T(\mathbf{k},\mu^2;x,\zeta)
%\end{equation}
%This implies in the double logarithmic approximation
%\begin{equation}\label{tmdcss2}
%\frac{\partial T(\mathbf{k},\mu^2;x,\zeta)}{\partial \ln\mu^2}=-\frac{\alpha_s N_c}{2\pi}\zeta\ T(\mathbf{k},\mu^2;x,\zeta)
%\end{equation}
%which is precisely the second CSS equation up to single logarithmic terms. We believe that in order to recover the single logarithmic terms one has to go beyond the leading order evolution in the sense discussed in \cite{dglapbfkl}. 

%Admittedly our derivation of \eqref{tmdcss2} is not quite satisfactory as it does not directly rely on the dynamics of the evolution. We could of course directly differentiate the definition \eqref{tmdmu} to obtain the left hand side of this equation. This should mean that within our evolution the TMD satisfies a nontrivial consistency condition that allows it to satisfy \eqref{tmdcss2}. This point is well worth understanding, but we will not pursue it further here.

To summarize, in order to properly understand effects related to the evolution of gluon TMD it is important to realize that its evolution is not given by the JIMWLK equation. Instead TMD evolves with $H_{CSS-JIMWLK}$, and only the initial condition for this evolution at $\zeta=\ln 1/x$ is generated by the standard JIMWLK evolution. This has to be consistently taken into account in processes where the relevant value of $x$ is set by the kinematics while one considers rapidity evolution beyond this value. We believe this must be important for the correct treatment of Sudakov emissions in processes like back to back dijet production in DIS \cite{dijet}.

%Now instead of introducing the dual charge rotation operators, one can of course write the kernel in terms of the matrices $U$. Formally this is easily done using \eqref{dj} in the penultimate expression in \eqref{nb}:
%\begin{equation}
%[N_\perp(b_L-b_R)]^a_i\simeq g^2\frac{\partial_i}{\partial^2}\left[(\partial D-D\partial)\frac{1}{D\partial}\mathcal{J}_R\right]^a
%\end{equation}
%Here $\mathcal{J}_R$ is the right rotation operator when acting on the standard Wilson line $U$.
%To get a more explicit expression we would need to find the inverse operator to $D\partial$.

\section*{Acknowledgement}
We thank Tolga Altinoluk, Nestor Armesto, Guillaume Beuf and Alina Czajka for many useful discussions related to the subject of this work. We are also grateful to Nestor Armesto and Florian Cougulic for comments on the draft.
The work of H.D. and A.K. is supported by the NSF Nuclear Theory grant 2208387.
This research    was supported  by   Binational Science Foundation  grant \#2022132.
 ML's work was funded by Binational Science Foundation grant \#2021789  and by the ISF grant \#910/23. 
 This research was also supported by
 MSCA RISE 823947 “Heavy ion collisions: collectivity and precision in saturation physics” (HIEIC).


\begin{thebibliography}{99}
\bibitem{footnote} Again, we refer the reader to \cite{ming} for the explanation of the relation of the ordering of the factors of $\hat j$ (and therefore $\hat b$)and the Wigner-Weyl representation of quantum mechanics of charges in terms of classical phase space variables.
\bibitem{footnote2} To be precise this expression is only valid in the dilute limit. For a dense projectile the TMD defined in \eqref{tmd} is not equal to the correlator of the classical fields, since the gluon state contains a squeezing factor. 
The full expression would contain also the factor $N_\perp$ as in \eqref{eq:Nperp_exp1} \cite{Altinoluk:2009jf}.
\bibitem{footnote3} More precisely, there is also a positive contribution to the gluon number due to splitting of gluons with even higher longitudinal momentum. In fact it turns out that in the evolution of $j_x$ the two effects almost completely cancel each other, so that the change in $j_x$ is very small. In the evolution of the TMD however, the loss strongly dominates over the gain leading to the Sudakov suppression factor.
\end{thebibliography}
\end{document}